# Model Independent Description of amplification and saturation using Green's Function

Yichao Jing*[1], Vladimir N. Litvinenko[1,2], Yue Hao[1], and Gang Wang[1]
[1] *Brookhaven National Laboratory, Upton, NY 11973, USA*
[2] *Department of Physics and Astronomy, Stony Brook University*

High-gain Free Electron Laser (FEL) is one of the many electron-beam instabilities that have a number of common features linking the shot noise, the amplification and the saturation. In this paper, we present a new, model-independent description of the interplay between these effects. We derive a simple formula for a maximum attainable gain before instability saturates. Application of this model-independent formula to FELs is compared with FEL theory and simulations. We apply the limitations resulting from these findings to FEL amplifiers used for seeded FELs and for Coherent electron Cooling.

## INTRODUCTION

FEL amplifiers can be used for a number of applications ranging from the High Gain Harmonic Generation (HGHG) [1] and the amplifying coherent FEL seeds [2] to the Coherent electron Cooling (CeC) [3]. Studies of the gain and the saturation in such FELs have captured wide interest and several criteria/estimations of saturation have been proposed [4, 5].

It is well known that e-beam instabilities, including those in FELs, can be described by a set of self-consistent Maxwell's and Vlasov's equations. In the classical limit, Maxwell's equations are linear while Vlasov's equation is not. Hence, the latter is responsible for the saturation. On the other hand, it is a well-established fact that Vlasov's equation can be linearized when the density modulation is significantly smaller compared with initial beam density. In other words, Vlasov's equation becomes nonlinear (which can cause the saturation of the instabilities) when the density modulation becomes comparable with the initial beam density $|\delta n| \sim n_o$.

When $|\delta n / n_o| << 1$ we can use linearized Vlasov's equation, which then can be represented by a Green's function. Here, for compactness, we consider a one-directional instability: see Appendix A for detailed derivation in the 3D case. The linear response of the one-dimensional system on a perturbation can be described by one-dimensional Green's function,

$$n(z,\tau) = n_o + \delta(z - z_o) + G_\tau(z - z_o), \tag{1}$$

where $n_0$ is the initial electron beam density, $\delta n = \delta(z - z_o)$ is a local perturbation at $z = z_o$ and $G_\tau(z - z_o)$ is the system's response on this perturbation at time τ. The Green's function satisfies natural conservation law:

* Electronic address: yjing@bnl.gov

$$\int_{-\infty}^{\infty} G_\tau(z)dz = 0\,. \tag{2}$$

Naturally this model is applicable for theoretically tractable 1D FELs. Furthermore, this 1D model is also a reasonable approximation for a 3D FEL when the diffraction smoothens transverse features while the longitudinal oscillations remain fast. We also consider a response of the system to be much shorter compared with the electron bunch length, e.g. the e-beam density could be considered locally constant.

Below we present detailed simulation study of the saturation of the Green's function using time resolved mode of FEL code Genesis 2.0 [6] for exact 3D FEL cases. We present the evolution and saturation of the Green's function (to be exact – the FEL response on a $\delta$-like initial perturbation) in FEL as function of longitudinal position. We compare the simulation results with the theoretical predictions for various 3D FELs operating in wide range of spectra: from the IR to the soft X-rays.

## GREEN'S FUNCTION AND ITS SATURATION

A 1D electron distribution can be described by its initial density:

$$n_o(z) = \sum_{i=1}^{N} \delta(z - z_i)\,, \tag{3}$$

where N is the number of electrons in the bunch. If initial distribution is random, eq. (3) is equivalent to the shot noise in the beam. In the absence of other perturbations (such as external sources of EM field or energy/velocity modulation in electron beam), eq. (3) fully describes the initial conditions. The linear response of the system then can be written as:

$$n(z,\tau) = \sum_{i=1}^{N} \delta(z - z_i) + \sum_{i=1}^{N} G_\tau(z - z_i)\,. \tag{4}$$

If the system has an initial external density perturbation $\delta n_o(z)$, the linear system response would be

$$\delta n(z,\tau) = \int_{-\infty}^{\infty} \delta n_o(\xi) G_\tau(z - \xi) d\xi\,. \tag{4a}$$

We have special interest in a specific can of external perturbations induced by ions co-propagating in CeC modulator, as shown in Fig. 1. As described in [3], each ion induces a point-like density modulation in electron beam with the total charge of *eX ~ -eZ*. It corresponds to a density perturbation approximately described by $X \cdot \delta(z - z_j)$, where $z_j$ is the location of the perturbation [1].

[1] To be exact, we should use a convolution of the induced density modulation by ions in the modulator with Green's function:

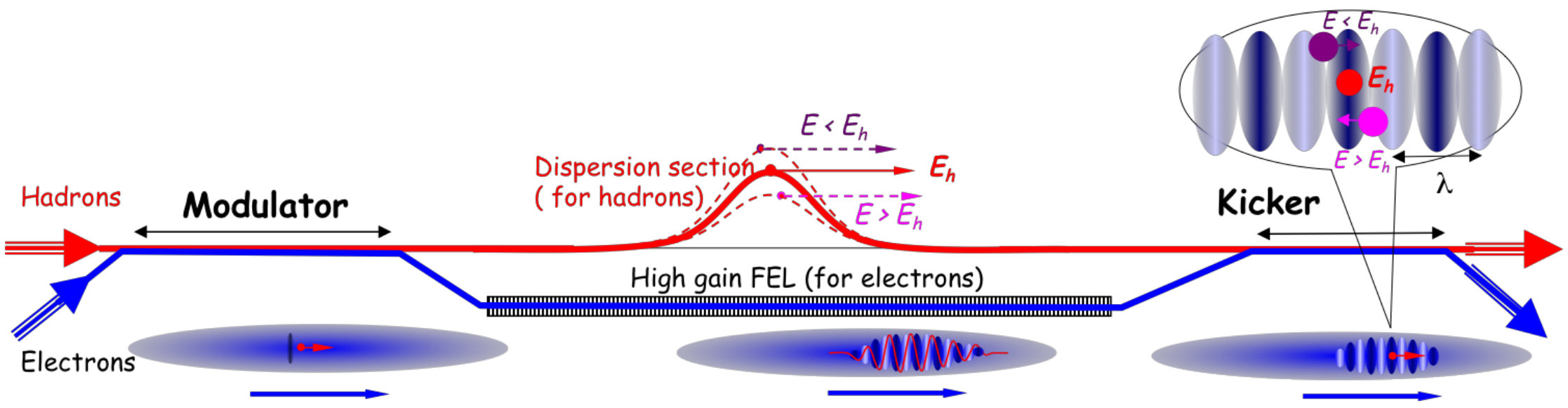

Fig. 1. A general schematic of the classical Coherent Electron Cooler [3] comprising three sections: A modulator, an FEL plus a dispersion section, and a kicker. For clarity, the size of the FEL wavelength, λ, is exaggerated grossly. In the CeC, the electron- and hadron-beams have the same velocity and co-propagate in the vacuum along a straight line in the modulator and the kicker. The CeC works as follows: In the modulator, each hadron (with charge, Ze, and atomic number, A) induces density modulation in electron beam that is amplified in the high-gain FEL; in the kicker, the hadrons interact with the beam's self-induced electric field and experience energy kicks toward their central energy. The process reduces the hadrons' energy spread, i.e., it cools the hadron beam.

Thus, the resulting density is described by:

$$n(z,\tau)=\sum_{i=1}^{N_e}\delta(z-z_i)+\sum_{i=1}^{N_e}G_\tau(z-z_i)+X\sum_{j=1}^{N_i}G_\tau(z-z_j). \qquad (5)$$

For the modulation of interest, i.e., within a wavelet $\{0,\lambda_o\equiv 2\pi/k_o\}$, we can calculate bunching factor, corresponding to the relative local density modulation:

$$b(\tau)=\frac{\int_0^{\lambda_o} n(\tau,z)e^{ik_o z}\,dz}{\int_0^{\lambda_o} n(\tau,z)\,dz}=\frac{\sum_{i,z_i\in\{0,\lambda_o\}} e^{ik_o z_i}+g(z_i)\sum_{i=1}^{N_e}e^{ik_o z_i}+Xg(z_j)\sum_{j=1}^{N_i}e^{ik_o z_j}}{M_e} \qquad (6)$$

where $M_e$ is the number of electrons in the wavelet $\{0,\lambda_o\}$ . We define the gain envelope of the instability through the integral of the Green's function within the wavelet,

$$\int_o^{\lambda_o} G_\tau(z-z_i)e^{ik_o z}\,dz=e^{ik_o z_i}\int_{-z_i}^{\lambda_o-z_i} G_\tau(z)e^{ik_o z}\,dz=e^{ik_o z_i}g(z_i), \qquad (7)$$

---

$$X\sum_{j=1}^{N_i}G_\tau(z-z_j)\rightarrow\sum_{j=1}^{N_i}G_{\tau h}(z-z_j);\quad G_{\tau h}(z-z_j)=\int G_{\tau h}(z-\zeta)\rho_h(\zeta-z_j)d\zeta\,.$$

where $\rho_h$ is a density modulation induced by the hadron in the CeC modulator [1]. For cases of interest, the details of $\rho_h$ are not important, when its duration is much shorter than that wavelength of the FEL amplifier.

where

$$g\left(z_i\right)=\int_{-z_i}^{\lambda_o-z_i} G_\tau(z)e^{ik_oz}\,dz. \tag{8}$$

Calculating the RMS value of the bunching factor, we assume the absence of correlation between electrons and hadrons, i.e., a random Poisson distribution of their initial phases:

$$\begin{gathered}\left\langle\left|M_e b(\tau)\right|^2\right\rangle=\left|\sum_{i,z_i\in\{0,\lambda_o\}} e^{ik_oz_i}+g\left(z_i\right)\sum_{i=1}^{N_e}e^{ik_oz_i}+X\left(z_j\right)g\left(z_j\right)\sum_{i=1}^{N_i}e^{ik_oz_i}\right|^2;\\ \left|\sum_{i,z_i\in\{0,\lambda_o\}} e^{ik_oz_i}+g\left(z_i\right)\sum_{i=1}^{N}e^{ik_oz_i}\right|^2=\sum_{i,z_i\in\{0,\lambda_o\}}\left(1+2\operatorname{Re}g\left(z_i\right)\right)+\sum_{i=1}^{N_e}\left|g\left(z_i\right)\right|^2+X^2\sum_{j}^{N_i}\left|g\left(z_j\right)\right|^2\end{gathered}. \tag{9}$$

In other words,

$$\begin{gathered}\left\langle\left|M_e b(\tau)\right|^2\right\rangle=M_e+2\sum_{i,z_i\in\{0,\lambda_o\}}\operatorname{Re}g\left(z_i\right)+\sum_{i=1}^{N_e}\left|g\left(z_i\right)\right|^2+\sum_{j}^{N_i}X^2\left(z_j\right)\left|g\left(z_j\right)\right|^2\\ \left\langle\left|M_e b(\tau)\right|^2\right\rangle=M_e(1+2\cdot\left\langle\operatorname{Re}g\left(z\right)\right\rangle_{z\in\{0,\lambda_o\}})+\int_{-\infty}^{\infty}\Lambda_e(z)\left|g\left(z\right)\right|^2dz+\int_{-\infty}^{\infty}\Lambda_I(z)X^2(z)\left|g\left(z\right)\right|^2dz\end{gathered} \tag{10}$$

where $\Lambda_e$ and $\Lambda_i$ are linear densities of electrons and ions respectively

$$\Lambda_e(z)=\left\langle\frac{\sum_{z_i\in\{z-\Delta z,z+\Delta z\}}1}{2\Delta z}\right\rangle_{\Delta z\to 0};\Lambda_I(z)=\left\langle\frac{\sum_{z_j\in\{z-\Delta z,z+\Delta z\}}1}{2\Delta z}\right\rangle_{\Delta z\to 0}.$$

For a continuous beam with fixed density, we have simple expression of

$$\Lambda_e=\frac{M_e}{\lambda_o};\Lambda_I=\frac{M_I}{\lambda_o},$$

giving us the following expression for the RMS value of bunching:

$$\left\langle\left|M_e b(\tau)\right|^2\right\rangle=M_e(1+2\cdot\left\langle\operatorname{Re}g\left(z\right)\right\rangle_{z\in\{0,\lambda_o\}})+\Lambda_e\int_{-\infty}^{\infty}\left|g\left(z\right)\right|^2dz+X^2\cdot\Lambda_I\int_{-\infty}^{\infty}\left|g\left(z\right)\right|^2dz \tag{11}$$

Thus, the modulation is determined by the effective correlation length, which is defined as:

$$\int_{-\infty}^{\infty}\left|g\left(z\right)\right|^2dz=g_{\max}^2N_c\lambda_o. \tag{12}$$

Our hypothesis is that the FEL is saturated at $|b|\to 1$, giving us an estimate for the maximum attainable gain:

$$1 + g_{\max}^2 N_c \left(1 + X^2 \cdot \frac{M_I}{M_e}\right) \le M_e \ , \tag{13}$$

or explicitly for $M_e >> 1$

$$g_{\max} \le \sqrt{\frac{M_e}{N_c}} \cdot \left\{ \begin{array}{c} 1, \ \ SASE \ \ FEL \\ \left(1 + X^2 \cdot \dfrac{M_I}{M_e}\right)^{-1/2}, CeC \end{array} \right\}, \tag{14}$$

where the second multiplier takes into account the shot noise at the FEL amplifier entrance. We can rewrite eq. (14) in practical units using e-beam's peak current $I_p$ and wavelength $\lambda_o$:

$$M_e = \frac{I_{pe}\lambda_o}{ec} = 2.08 \cdot 10^4 \cdot I_p[A]\lambda_o[\mu m] \tag{15}$$

Thus the estimated maximum attainable gain becomes

$$g_{\max} \le 144 \cdot \sqrt{\frac{I_{pe}[A] \cdot \lambda_o[\mu m]}{N_c \left(1 + \dfrac{X^2}{Z} \cdot \dfrac{I_{pI}}{I_{pe}}\right)}} . \tag{16}$$

In the absence of the shot noise induced by the ion bunch, e.g. in case of SASE FEL or other e-beam instability, it simplifies to

$$g_{\max} \le 144 \cdot \sqrt{\frac{I_{pe}[A] \cdot \lambda_o[\mu m]}{N_c}} \ . \tag{17}$$

Thus, we arrive to a very simple and neat formula for the maximum attainable gain where only the electron's peak current $I_p$, the FEL wavelength $\lambda_0$ and coherence length (measured in number of instability wavelength, $\lambda_0$) $N_c$ are involved. The result does not require the knowledge of the type of FEL or other e-beam instability. It also does not involve other properties of the electron beam. As we will see below, it is also suitable for 3D-FEL . However, there is no analytical expression for $N_c$ for an arbitrary 3D FEL.

In the following section, we present result of simulating a statistically representative set of initial conditions at the FEL entrance. We used 3D FEL code Genesis for a number of random shot noise sets. We simulated the evolution of the FEL "with" and "without" a small the δ-function-like perturbation. We extract the Green's function's amplitude and phase evolution by subtracting the bunching factors for case "without" from that "with" the δ-function-like perturbation. Note that bunching factor is a complex number with the amplitude and the phase. Hence, the Green's function is also complex. We calculate the average values of the Green's functions for all sets as well as RMS values of its amplitude and phase variations. Naturally, in linear regime the Green's function does not depend on the noise set. Hence, we evaluated its saturation by observing the fluctuation its phase and amplitude.

We repeated this process for a number of FEL wavelengths and summarize the results in Table. I.

## NUMERICAL FEL AMPLIFIER ANALYSIS

As mentioned above, we consider a response of the system (coherence length) to be much shorter compared with the electron bunch length. Thus we are able to slice the electron bunch into wavelets (with step of wavelength $\lambda_0$), where each slice is represented by local bunching factor (with amplitude and phase). Genesis is a perfect tool to simulate the evolution of the bunching factor in a FEL.

We run Genesis in time-resolved mode to simulate the FEL amplification and saturation process. To extract the information about the Green's function, a reaction on $\delta$-function like perturbation, for each shot noise set, we generated two electron bunch distributions. The first distribution has a random Poison shot noise, generated by Genesis. This is a typical and well-tested setting for SASE FEL simulations. The second distribution is identical to the first one with exception of a single wavelet where we added a small fixed bunching factor in a single wavelet located at the middle of the bunch. Technically, we use quiet start to generate one set slice with initial bunching factor at the level of interest, e.g. $10^{-4}$ and superimpose this slice on the middle of the first distribution. Thereafter we propagated these two bunches (one only with the shot noise, the other, with the shot noise and the perturbation) through the FEL to record evolution of the bunching factors in each wavelet (amplitude and phase) as function of longitudinal position in the wiggler.

$$b_1(z) = |b_1(z)| e^{i\theta_1(z)}, b_2(z) = |b_2(z)| e^{i\theta_2(z)} \tag{18}$$

The bunching caused by the perturbation is the difference of these two results and can then be written as:

$$b_s(z) = |b_s(z)| e^{i\theta_s(z)} \equiv b_2(z) - b_1(z), \tag{19}$$

or specifically

$$\begin{gathered} |b_s(z)| = |b_1(z) - b_2(z)|; \\ \theta_s(z) = \arctan 2\left(|b_1(z)|\cos\theta_1 - |b_2(z)|\cos\theta_2, |b_1(z)|\sin\theta_1 - |b_2(z)|\sin\theta_2\right). \end{gathered} \tag{20}$$

The *arctan2* function gives correct phase values within the range of [-π, π]. However it generates an artificial "jump" from –π to π when the phase crosses the boundary. This could produce superficial discontinuities of the bunching phase (modular of 2π). We monitor this and give the correction (modular of 2π) whenever this "jump" would occur.

Simulating FEL process in Genesis using the entire electron bunch would require a very large number of slices and even larger number of macro-particles. Number of sliced would be equal to the total bunch length (~ 1cm) divided by the FEL wavelength (from μm to nm), e.g. about $10^4$-$10^7$ slices. Each slice should have a sufficiently large number of macro particles for proper FEL statistics. It would bring the number of macro-particles to $10^8$-$10^{11}$. Thus, the full bunch FEL simulation would be very time consuming and, in fact, unnecessary. In FEL the information propagates (slips) for one wavelength per wiggler period. Hence, the number of slices needed for extracting the response on the initial perturbation (Green's function) is equal to that of the number of the wiggler periods, which is only few hundreds. Naturally, we added some buffer to remove the influence of the boundaries. Therefore, we used 1000 slices with 16384 macro particles per slice. For statistics we use 32 random shot noise sets in Genesis. It allows us to study where the response of the initial perturbation saturates.

We simulated a number of FEL amplifiers operating in various spectral ranges: from infrared (planned for CeC proof of principle test), visible (planned to be used for eRHIC CeC), VUV (a candidate for a LHC CeC), and hard X-ray (LCLS). Detailed parameters are listed in Table I.

TABLE I: Parameters for FEL simulation

| FEL type<br>Parameters | Infrared<br>PoP CeC | Visible<br>eRHIC CeC | VUV<br>LHC CeC | Hard X-rays<br>LCLS |
|---|---|---|---|---|
| Beam energy (MeV) | 21.8 | 136 | 3812.3 | 13643.7 |
| Beam current (peak, A) | 100 | 10 | 30 | 3400 |
| Normalized emittance (μm rad) | 5 | 1 | 1 | 1.2 |
| Momentum spread ($\sigma_p/p$) | $1\times10^{-3}$ | $1.5\times10^{-5}$ | $2.5\times10^{-5}$ | $1.05\times10^{-4}$ |
| Undulator period (cm) | 4 | 3 | 10 | 3 |
| Undulator strength, $a_w$ | 0.4 | 1 | 10 | 2.4756 |
| Radiation wavelength | 12.7 μm | 423.5 nm | 90.7 nm | 0.15 nm |
| $N_c$ | 35.8 | 102 | 70.6 | 14.5[a] |

[a] The temporal coherence mode number of LCLS is taken from the measured value shown in [7].

While we studied each of the four cases in details, here, for conciseness, we will describe one case: the visible FEL amplifier we plan to use for eRHIC CeC.

First, we tested that a single slice perturbation with bunching at $10^{-4}$ does not saturate in our FEL amplifier. Second, we checked that its addition also does not affect the saturation of the SASE (short-noise) signal. Then, we used 32 random shot noise sets to extract the evolution in the FEL including the bunching Green's function as in eq. (20). Fig.2 shows the evolution of the average bunching amplitude (see eq. (18)) and the maximum of the response function shown in eq. (19).

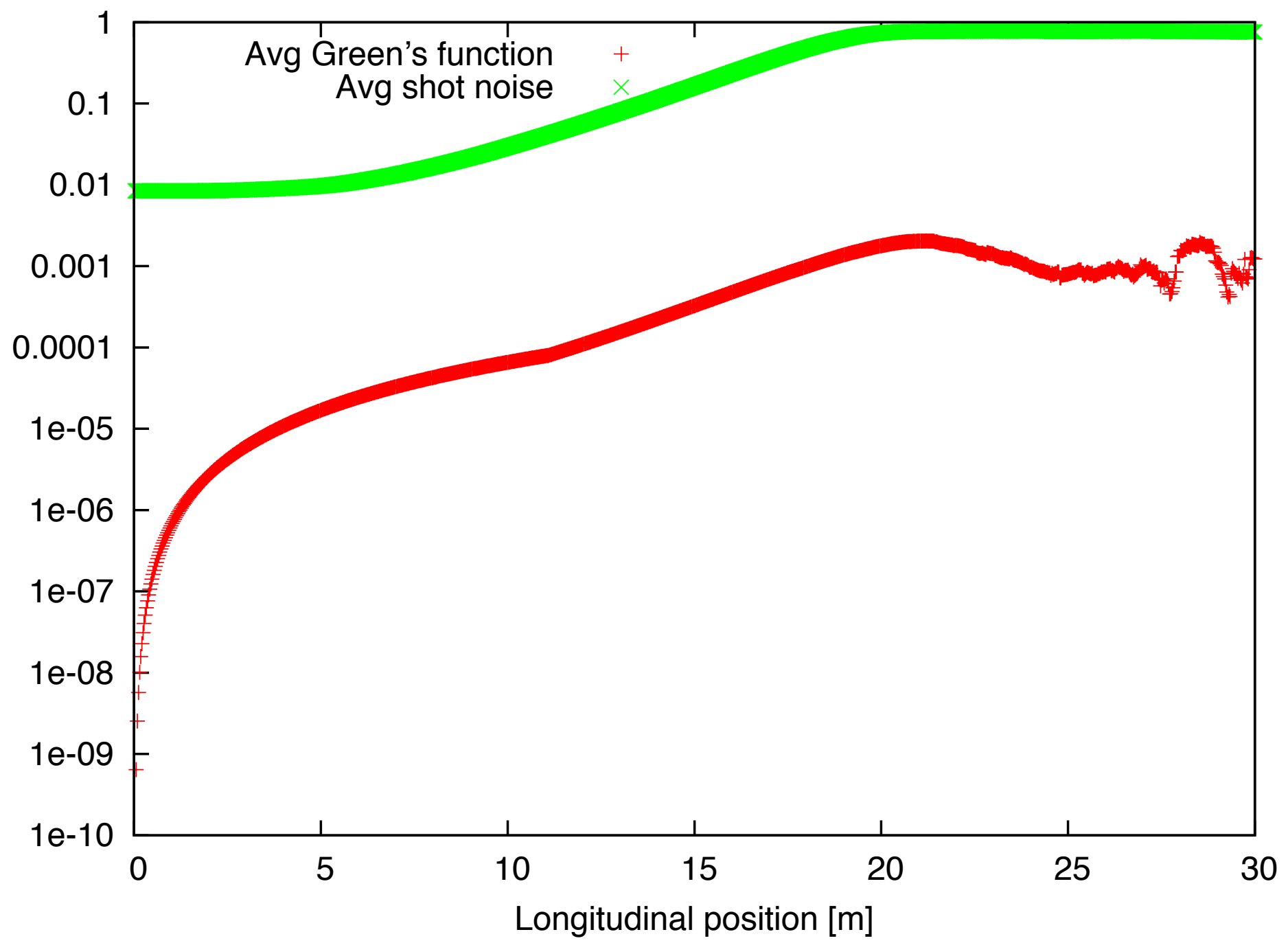

FIG. 2: Evolution of the amplitude of the bunching factors in eRHIC CeC FEL amplifier averaged over 32 random shot noise sets. Green curve is the evolution of the bunch average amplitude of the bunching factor. Red curve shows the evolution of the peak amplitude of the bunching Green's function (without initial perturbation) defined in eq. (4).

The evolution of the average bunching factor (Green's curve) is typical to SASE FELs, with initial formation period, followed by the exponential growth and, finally, saturation with $|b_{SASE}| \sim 1$. The evolution of peak of the Green's function is rather different. It has initial range of slight decline, followed by exponential growth and saturation (destruction). The maximum amplification of the initial perturbation is reached just before the FEL saturation (at approximately 20 m) and reaches 27.3.

Naturally, the Green's function cannot be simply described by the amplitude of its peak value: there are also some other important features as the location of its peak and the phase of the bunching modulation at that location. Fig. 3 illustrates the latter. The phase of the response function has two type of behavior: smooth evolution till about 20 m into the FEL, and jumpy piece-vice behavior after that. There is a slight phase jump at about 12 m into the FEL, which represents the slippage of the longitudinal slice that carries the maximum bunching information. As we see earlier in Fig. 2, this slice, carrying the response (the Green's function) from the initial perturbation (at zero position), grows exponentially in FEL while slips through the longitudinal positions. Further more, this change of slices does not have dependence on the random shot noise as shown in Fig. 3. Hence, we are mostly interested in the evolution from this point further, e.g. gain larger than one (which can also be seen from Fig. 2 where the bunching factor grows larger than the initial perturbation 1e-4). The evolution before this point mostly represents academic interest (see below). Fig. 3 also shows the RMS spread of the phase at the peak value which is deep saturation reaches about 2 radian. The latter means that the phase information about initial perturbation is washed away by the random noise. For CEC the preservation of the phase information is of great importance, hence RMS phase noise about 1 radian are not acceptable. It means that starting from about 22 m, the information is washed away.

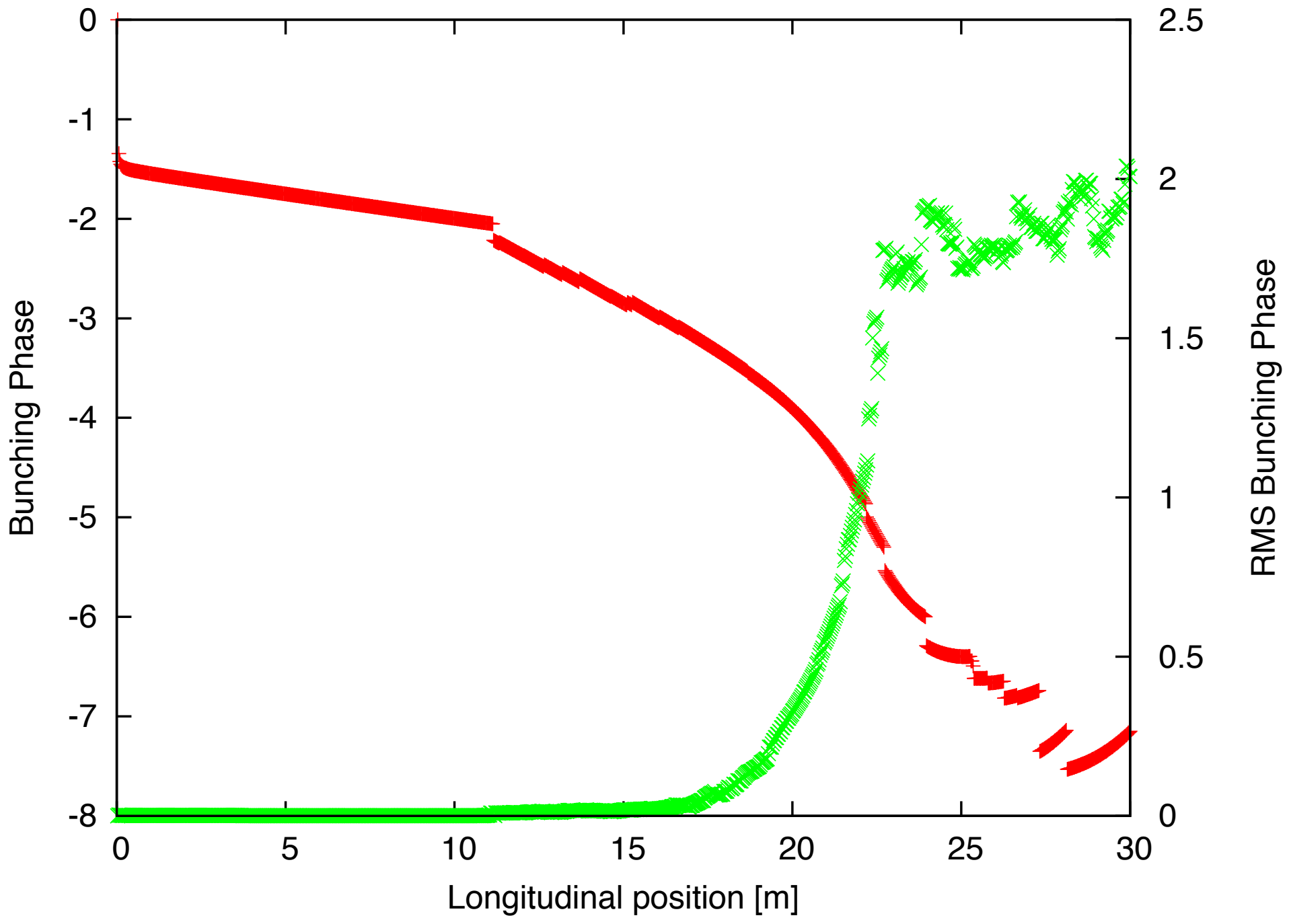

FIG.3: Plots of the Green's function's (eq. (19)) phase as at its peak value as function of position in the FEL wiggler. 32 random shot noise sets are used to generate these data. Red curve shows the average phase and Green's curve shows RMS variation of this phase.

The entire process – the slippage, amplification and saturation of the FEL response – can be visualized by the evolution of the response's profile within the FEL. Figure. 4 where the evolution of the bunching induced by initial perturbation is shown as a function of the slippage (in unit of radiation wavelength) at several locations in the wiggler. The initial single-wavelet perturbation with bunching amplitude of $10^{-4}$ and zero phase is placed at the middle of the bunch having natural random shot noise. Since we simulate the bunch with the shot noise, we are subtracting its contribution. The only non-trivial part is coming from nonlinear interaction between the shot noise and the coherent signal induced by the perturbation.

This slippage of the Green's function can thus be easily visualized in Fig. 4. The bunching envelope starts as localized at where the initial perturbation locates then starts to develop to the entire bunch. At around 12 m, the peak of the envelope starts to grow higher than initial perturbation at 1e-4 (gain larger than 1). We can also see that when it approaches saturation, the envelope starts to wiggle which indicates the random shot noise starts to play a more important role.

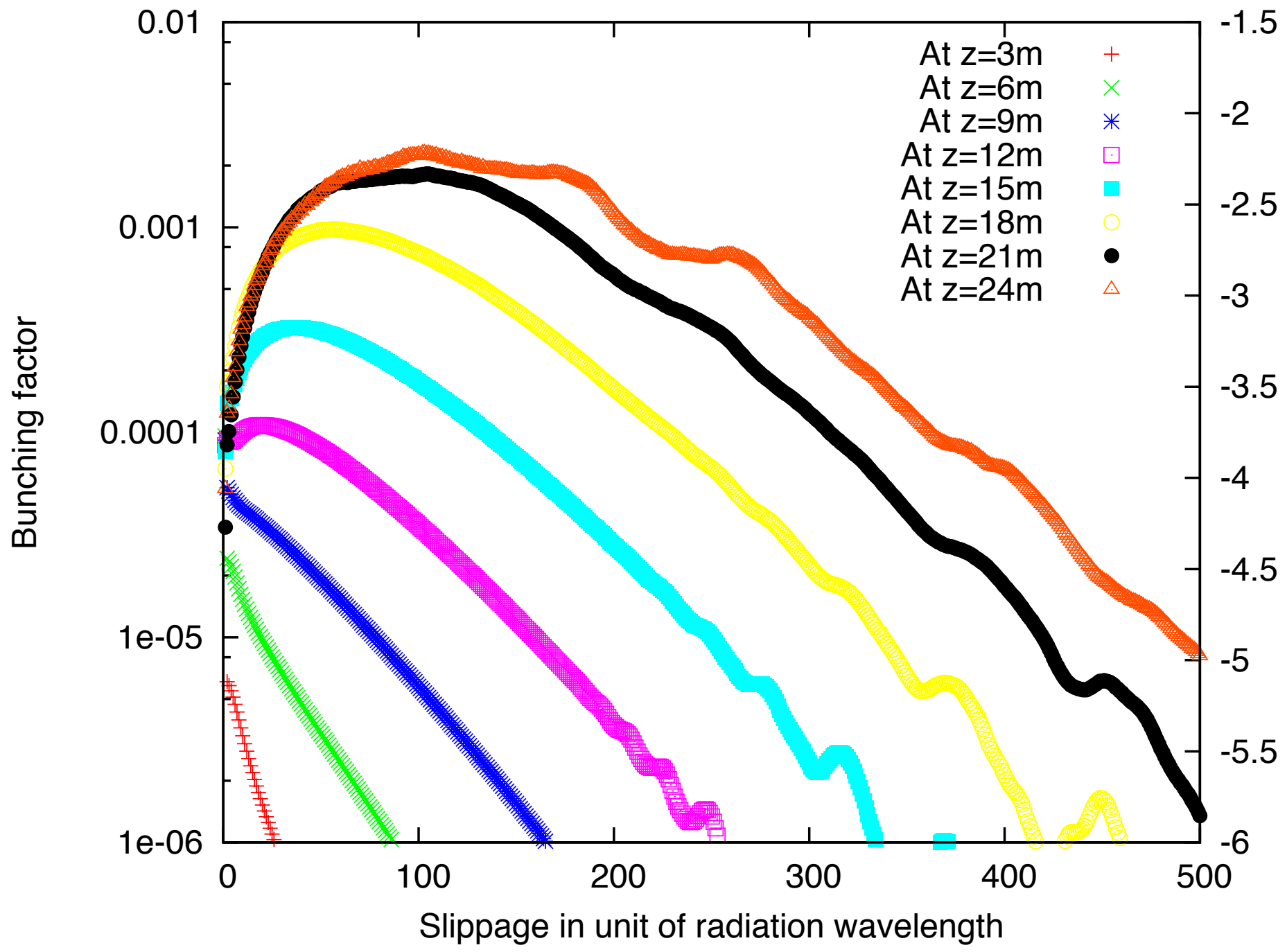

FIG.4: Plot of the evolution of the bunching induced by the initial perturbation as function of the slippage of a number of locations in the wiggler with shot noise.

We calculate the mean and fluctuation of this envelope over random shot noises and select several characteristic locations in the wiggler to present our results in Fig. 5.

(a) (b)

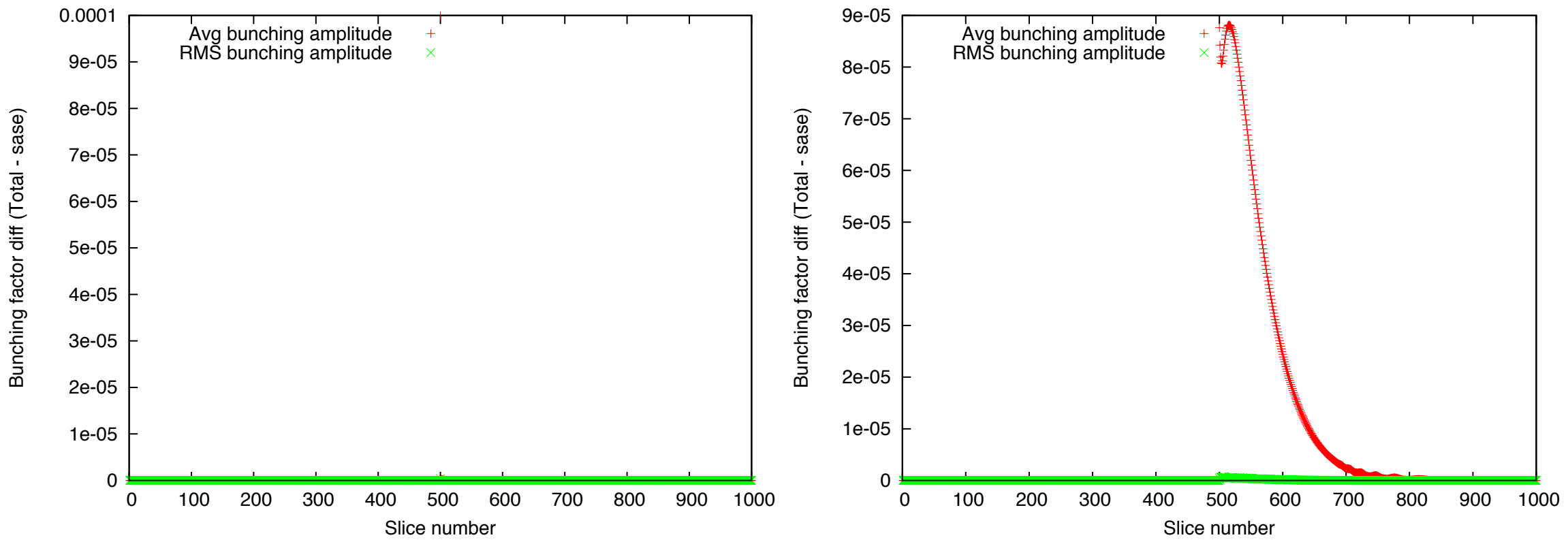

(c) (d)

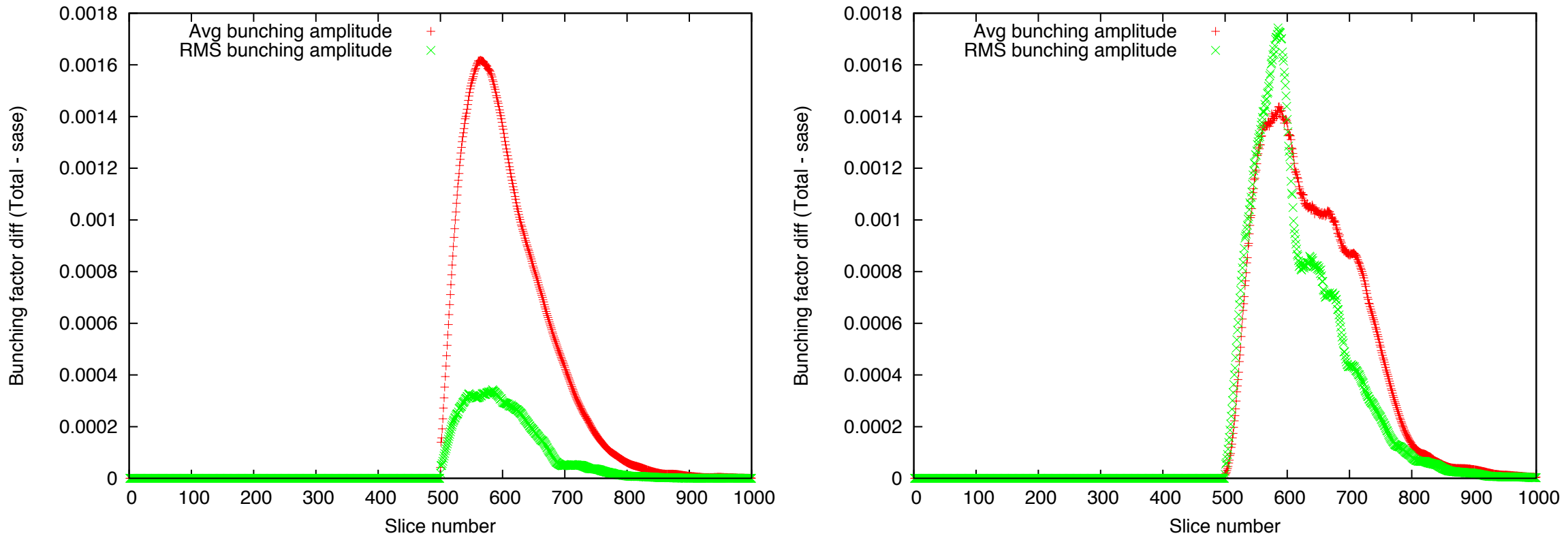

FIG. 5: The bunching as function of the wavelet (slice) number at four location in the FEL wiggler: (a) near the entrance (z = 0.6m); (b) at z=11.4 m (where amplitude of the response is about to overtake that of the initial perturbation; (c) just at the on-set of the saturation, at z=19.8 m and (d) at z=22.0 m, well in the saturation. We used 32 random shot noise sets. Red curves indicate for average value of the amplitude, and Green's curves show the RMS spread of the bunching amplitudes.

Figure. 5(a) shows the initial single-wavelet perturbation itself. We note that the initial density perturbation travels with the speed of the electron beam, while the response moves with a slightly higher velocity, e.g. the group velocity of the FEL. When the electrons propagate in the FEL wiggler, three wave-packets develop from the initial perturbation: the density, the energy modulation and the EM radiation. The radiation wave-packet slips forward of one FEL wavelength per FEL wiggler period. After the initial process of the correlations build-up between three wave-packets, the exponential growth starts. At some point the response overtakes the initial perturbation – this moment is illustrated in Fig. 5 (b).

The maximum of bunching amplitude is reached at the on-set of the saturation. Fig. 5 (c) shows the FEL response at this location, which has a FWHM length about 150 FEL wavelengths.

Finally, as show in Fig. 5(d), the response would saturate, which both amplitude and phase randomized by nonlinear interaction with the shot noise of the beam. Naturally, this is no longer useful for amplification of desirable signal.

## DISCUSSION

We did similar analysis to the above for each of four FELs listed in Table I and found maximum non-saturated gain for each of them. Table II shows the comparison of these values with a model-independent estimate given by eq. (17). For a simple 1D estimate derived without taking into account of many important 3D effects, the agreement is remarkably good.

Table II: Comparison of bunching gain of theory and GENESIS simulation

| Name | $g_{max}$, eq. (17) | $g_{max}$, Genesis simulations |
|---|---|---|
| Infrared FEL | 857.5 | 777 |
| Visible FEL | 29.3 | 27 |
| VUV FEL | 28.3 | 18.7 |
| Hard X-ray FEL | 27 | 21.1 |

To avoid a possibility of these agreement being a lucky choice of the random shot noise sets, we repeated these simulation for a completely different sets of random noise. The findings were identical.

## CONCLUSIONS AND ACKNOWLEDGEMENTS

In this paper, we derive a simple model-independent formula estimating the maximum attainable gain for an electron beam instability (such as FEL) in the presence of the shot noise. We compared it with direct 3D FEL simulations for a variety of wavelengths using code Genesis in the time resolved mode. We found a very good agreement between the simulation and our theoretical estimate.

This work is supported by the U.S. Department of Energy under Contract No. DEAC02-98CH10886 and by the NSF under Grant PHY-1415252.

**APPENDIX**

The phase space distribution of an electron beam can be expressed into a sum of delta functions, i.e.

$$\tilde{f}(\vec{x},\vec{p},t)=\sum_{i=1}^{N_e}\delta(\vec{x}-\vec{x}_i(t))\delta(\vec{p}-\vec{p}_i(t)). \quad \text{(A.1)}$$

Let the expectation value of the distribution function after ensemble average be $\bar{f}(\vec{x},\vec{p},t)$, which is usually a smooth function. At the starting time, $t=t_0$, each 6-D delta function in the sum of eq. (A.1) presents a perturbation to the ensemble averaged initial distribution, $\bar{f}(\vec{x},\vec{p},t_0)$. We describe the electron density perturbation induced by a point-like perturbation by the Green's function, i.e.

$$\delta(\vec{x}-\vec{x}_i(t_0))\delta(\vec{p}-\vec{p}_i(t_0))\to\delta(\vec{x}-\vec{X}_i(t))\delta(\vec{p}-\vec{P}_i(t))+G(\vec{x},\vec{x}_i(t_0),\vec{p},\vec{p}_i(t_0),t-t_0), \quad \text{(A.2)}$$

where $\vec{X}_i(t)$ and $\vec{P}_i(t)$ are the trajectories of the electron in the absence of the interaction that causing instability and hence the unperturbed distribution is given by

$$\tilde{f}_0(\vec{x},\vec{p},t)=\sum_{i=1}^{N_e}\delta(\vec{x}-\vec{X}_i(t))\delta(\vec{p}-\vec{P}_i(t)). \quad \text{(A.3)}$$

Thus the phase space distribution at a later time $t>t_0$ is given by

$$\begin{aligned}\tilde{f}(\vec{x},\vec{p},t)&=\sum_{i=1}^{N_e}\delta(\vec{x}-\vec{x}_i(t))\delta(\vec{p}-\vec{p}_i(t))\\&=\tilde{f}_0(\vec{x},\vec{p},t)+\sum_{i=1}^{N_e}G(\vec{x},\vec{x}_i(t_0),\vec{p},\vec{p}_i(t_0),t-t_0)\end{aligned}. \quad \text{(A.4)}$$

Integrating eq. (A.4) over the whole phase space volume leads to

$$\sum_{i=1}^{N_e}\int_{-\infty}^{\infty}G(\vec{x},\vec{x}_i(t_0),\vec{p},\vec{p}_i(t_0),t-t_0)d^3xd^3p=0. \quad \text{(A.5)}$$

Assuming there is no correlation between any two electrons, eq. (A.5) reduces to

$$\int_{-\infty}^{\infty}G(\vec{x},\vec{x}_i(t_0),\vec{p},\vec{p}_i(t_0),t-t_0)d^3xd^3p=0\ , \quad \text{(A.6)}$$

for any $x_i(t_0)$ and $\vec{p}_i(t_0)$. According to eq. (A.4), the electron phase space density perturbation due to the instability can be written as

$$\delta\tilde{f}(\vec{x},\vec{p},t)\equiv\tilde{f}(\vec{x},\vec{p},t)-\tilde{f}_0(\vec{x},\vec{p},t)=\sum_{i=1}^{N_e}G(\vec{x},\vec{x}_i(t_0),\vec{p},\vec{p}_i(t_0),t-t_0)\ . \quad \text{(A.7)}$$

Requiring the RMS perturbation amplitude being smaller than the unperturbed phase space density leads to

$$\left\langle \delta\tilde{f}\left(\vec{x},\vec{p},t\right)^2\right\rangle = \left\langle \left[\sum_{i=1}^{N_e} G\left(\vec{x},\vec{x}_i\left(t_0\right),\vec{p},\vec{p}_i\left(t_0\right),t-t_0\right)\right]^2\right\rangle$$

$$= \left\langle \left[\sum_{i,j=1}^{N_e} G\left(\Gamma,\Gamma_i\left(t_0\right),t-t_0\right)\right]^2\right\rangle \tag{A.8}$$

$$< \left\langle \tilde{f}_0\left(\vec{x},\vec{p},t\right)\right\rangle^2$$

with $\Gamma$ representing a 6D phase space vector and the angle brackets means ensemble average. The summation in eq. (A.8) can be written as

$$\left[\sum_{i,j=1}^{N_e} G\left(\Gamma,\Gamma_i\left(t_0\right),t-t_0\right)\right]^2$$

$$= \left[\int_{-\infty}^{\infty} G\left(\Gamma,\xi,t-t_0\right)\tilde{f}\left(\xi,t_0\right)d^6\xi\right]^2$$

$$= \left[\int_{-\infty}^{\infty} G\left(\Gamma,\xi,t-t_0\right)\overline{f}\left(\xi,t_0\right)d^6\xi + \int_{-\infty}^{\infty} G\left(\Gamma,\xi,t-t_0\right)\delta\tilde{f}\left(\xi,t_0\right)d^6\xi\right]^2 . \tag{A.9}$$

$$= \left[\int_{-\infty}^{\infty} G\left(\Gamma,\xi,t-t_0\right)\overline{f}\left(\xi,t_0\right)d^6\xi\right]^2 + \left[\int_{-\infty}^{\infty} G\left(\Gamma,\xi,t-t_0\right)\delta\tilde{f}\left(\xi,t_0\right)d^6\xi\right]^2$$

$$+2\int_{-\infty}^{\infty} G\left(\Gamma,\xi,t-t_0\right)\overline{f}\left(\xi,t_0\right)d^6\xi \int_{-\infty}^{\infty} G\left(\Gamma,\xi,t-t_0\right)\delta\tilde{f}\left(\xi,t_0\right)d^6\xi$$

Taking the ensemble average of eq. (A.6) yields

$$\left\langle \left[\sum_{i,j=1}^{N_e} G\left(\Gamma,\Gamma_i\left(t_0\right),t-t_0\right)\right]^2\right\rangle$$

$$= \left[\int_{-\infty}^{\infty} G\left(\Gamma,\xi,t-t_0\right)\overline{f}\left(\xi,t_0\right)d^6\xi\right]^2 + \left\langle \left[\int_{-\infty}^{\infty} G\left(\Gamma,\xi,t-t_0\right)\delta\tilde{f}\left(\xi,t_0\right)d^6\xi\right]^2\right\rangle . \tag{A.10}$$

Writing the integral inside the second term on the right hand side of eq. (A.10) as a sum over phase space volume, i.e.

$$\int_{-\infty}^{\infty} G\left(\Gamma,\xi,t-t_0\right)\delta\tilde{f}\left(\xi,t_0\right)d^6\xi = \sum_{i=1}^{\infty} G\left(\Gamma,\xi_i,t-t_0\right)\delta\tilde{f}\left(\xi_i,t_0\right)\Delta\xi , \tag{A.11}$$

with $\Delta\xi$ being an infinitesimal phase space volume, the second term in the right hand side of eq. (A.10)can be written as

$$\left\langle \left[ \int_{-\infty}^{\infty} G(\Gamma,\xi,t-t_0)\delta\tilde{f}(\xi,t_0)d^6\xi \right]^2 \right\rangle$$

$$= \left\langle \sum_{i,j=1}^{\infty} G(\Gamma,\xi_i,t-t_0)G(\Gamma,\xi_j,t-t_0)\delta\tilde{f}(\xi_j,t_0)\delta\tilde{f}(\xi_i,t_0)\Delta\xi\Delta\xi \right\rangle . \qquad \text{(A.12)}$$

$$= \sum_{i,j=1}^{\infty} G(\Gamma,\xi_i,t-t_0)G(\Gamma,\xi_j,t-t_0)\left\langle \delta\tilde{n}(\xi_j,t_0)\delta\tilde{n}(\xi_i,t_0)\right\rangle$$

Since there is no correlation between two phase space volume at different locations, the following relation holds,

$$\left\langle \delta\tilde{n}(\xi_j,t_0)\delta\tilde{n}(\xi_i,t_0)\right\rangle = \left\langle \delta\tilde{n}(\xi_i,t_0)^2\right\rangle \delta_{i,j} = \left\langle \tilde{n}(\xi_i,t_0)\right\rangle \delta_{i,j} , \qquad \text{(A.13)}$$

and hence eq. (A.12)becomes

$$\left\langle \left[ \int_{-\infty}^{\infty} G(\Gamma,\xi,t-t_0)\delta\tilde{f}(\xi,t_0)d^6\xi \right]^2 \right\rangle = \sum_{i=1}^{\infty} G(\Gamma,\xi_i,t-t_0)^2 \left\langle \tilde{n}(\xi_i,t_0)\right\rangle$$

$$= \sum_{i=1}^{\infty} G(\Gamma,\xi_i,t-t_0)^2 \bar{f}(\xi_i,t_0)\Delta\xi \quad , \qquad \text{(A.14)}$$

$$= \int_{-\infty}^{\infty} \left[G(\Gamma,\xi,t-t_0)\right]^2 \bar{f}(\xi,t_0)d^6\xi$$

where we used the law of rare event (Poisson's distribution) in obtaining the second equation of eq. (A.13). Making use of eq. (A.10) and (A.14), the ensemble average of eq. (A.8) leads to

$$\left[ \int_{-\infty}^{\infty} G(\vec{x},\vec{x}_0,\vec{p},\vec{p}_0,t-t_0)\bar{f}(\vec{x}_0,\vec{p}_0,t_0)d^3x_0d^3p_0 \right]^2$$

$$+ \int_{-\infty}^{\infty} \left[G(\vec{x},\vec{x}_0,\vec{p},\vec{p}_0,t-t_0)\right]^2 \bar{f}(\vec{x}_0,\vec{p}_0,t_0)d^3x_0d^3p_0 < \bar{f}_0(\vec{x},\vec{p},t)^2 . \qquad \text{(A.15)}$$

For a uniform electron distribution, the first term on the right hand side of eq. (A.15) vanishes due to eq. (A.6), and we arrive at

$$\int_{-\infty}^{\infty} \left[G(\vec{x},\vec{x}_0,\vec{p},\vec{p}_0,t-t_0)\right]^2 d^3x_0d^3p_0 < \bar{f}_0(\vec{x},\vec{p},t) . \qquad \text{(A.16)}$$

Defining a characteristic phase space volume as

$$\Gamma_c \equiv \frac{\int_{-\infty}^{\infty} \left[G(\vec{x},\vec{x}_0,\vec{p},\vec{p}_0,t-t_0)\right]^2 d^3x_0d^3p_0}{\left|G(t-t_0)\right|_{\max}^2} , \qquad \text{(A.17)}$$

eq. (A.16) can be re-written as

$$\left|G(t-t_0)\right|_{\max} < \sqrt{\frac{\bar{f}_0(\vec{x},\vec{p},t)}{\Gamma_c}} \ . \qquad \text{(A.18)}$$

Eq. (A.18) suggests that the maximal amplitude of the Green's function is proportional to the square root of the phase space electron density and inversely proportional to the square root of the coherence volume.

Assuming $G(\vec{x},\vec{x}_0,\vec{p},\vec{p}_0,t-t_0)$ is much wider than the beam in all 5 dimensions except for the longitudinal dimension, we can approximate it by a constant over the beam occupied area i.e.

$$G(\vec{x},\vec{x}_0,\vec{p},\vec{p}_0,t-t_0) \to G(\vec{x}_{b\perp},\vec{x}_{0\perp},z,z_0,\vec{p}_{b\perp},\vec{p}_0,t-t_0) \equiv G_b(z,z_0,t-t_0) \ , \qquad \text{(A.19)}$$

where $\vec{x}_{b\perp}$ and $\vec{p}_{b\perp}$ are average coordinates of electrons at location $z$. Using eq. (A.19), we can reduce eq. (A.15) to

$$\left[\int_{-\infty}^{\infty} G_b(z,z_0,t-t_0)\bar{\rho}(z_0,t_0)dz_0\right]^2 + \int_{-\infty}^{\infty}\left[G_b(z,z_0,t-t_0)\right]^2 \bar{\rho}(z_0,t_0)dz_0 < \bar{f}_0(\vec{x},\vec{p},t)^2 . \qquad \text{(A.20)}$$

Assuming $\bar{\rho}(z_0,t_0)$ is uniform and using eq. (A.6) and (A.19), the first term in eq. (A.20) is

$$\left[\bar{\rho}(t_0)^2 \int_{-\infty}^{\infty} G_b(z,z_0,t-t_0)dz_0\right]^2 \approx 0 \, , \qquad \text{(A.21)}$$

and we arrive at the criteria for a directional instability:

$$\int_{-\infty}^{\infty}\left[G_b(z,z_0,t-t_0)\right]^2 \bar{\rho}(z_0,t_0)dz_0 < \bar{f}_0(\vec{x},\vec{p},t)^2 \ . \qquad \text{(A.22)}$$

Taking square root of both sides of eq. (A.22) and integrating over all the dimensions except for the longitudinal dimension yields

$$\Gamma_5\sqrt{\int_{-\infty}^{\infty}\left[G_b(z,z_0,t-t_0)\right]^2 \bar{\rho}(z_0,t_0)dz_0} < \bar{\rho}_0(\vec{x},\vec{p},t) \ . \qquad \text{(A.23)}$$

Taking square of both sides of eq. (A.23) leads to

$$\int_{-\infty}^{\infty} G_{1d}(z,z_0,t-t_0)^2 \bar{\rho}(z_0,t_0)dz_0 < \bar{\rho}_0(\vec{x},\vec{p},t)^2 \ , \qquad \text{(A.24)}$$

where we have defined the 1D Green's function as

$$G_{1d}(z,z_0,t-t_0) \equiv G_b(z,z_0,t-t_0)\Gamma_5 = G_b(x_{b\perp},x_{0\perp},\vec{p},\vec{p}_0,z,z_0,t-t_0)\Gamma_5 . \qquad \text{(A.25)}$$